\documentclass[cameraready]{Interspeech}

\title{An Ultra-Low-Bitrate Neural Speech Codec with Plain-to-Pseudo Synergistic Vector Quantization}

\author[affiliation={1}]{Xiao-Hang}{Jiang}
\author[affiliation={1}, correspondingauthor]{Yang}{Ai}
\author[affiliation={1}]{Fei}{Liu}
\author[affiliation={1}]{Rui-Chen}{Zheng}
\author[affiliation={2}]{Jian-Qing}{Gao}
\author[affiliation={1}]{Zhen-Hua}{Ling}
\author[affiliation={3}]{Ji}{Wu}

\address{
    $^1$ University of Science and Technology of China, China \\
    $^2$ iFLYTEK Co., Ltd., China, $^3$ Tsinghua University, China
}
\email{
\{jiang\_xiaohang, fliu215, zhengruichen\}@mail.ustc.edu.cn, \\
\{yangai, zhling\}@ustc.edu.cn, jqgao@iflytek.com, wuji\_ee@tsinghua.edu.cn
}
\keywords{neural speech codec, plain-to-pseudo synergistic vector quantizer, token prediction, ultra-low bitrate}

\usepackage{comment}
\usepackage{multirow}

\begin{document}

\maketitle
\addtolength{\textfloatsep}{-0.5cm}
\addtolength{\dbltextfloatsep}{-0.5cm}

\begin{abstract}
Most neural speech codecs use residual vector quantization (RVQ), in which later VQs contribute less but consume the same bitrate, leading to inefficiency. We propose P2PSynCodec, an ultra-low-bitrate neural speech codec with a plain-to-pseudo synergistic vector quantizer (P2PSVQ). P2PSVQ consists of one plain VQ and multiple pseudo VQs. The plain VQ produces basic tokens by quantization, while the pseudo VQs generate auxiliary tokens by neural prediction and incur zero transmitted bitrate. Thus, speech is decoded from the plain-VQ tokens together with predicted pseudo-VQ tokens, greatly reducing bitrate. Experiments show that P2PSynCodec achieves speech reconstruction quality comparable to competing codecs at 2.0 kbps while operating at only 0.5 kbps, demonstrating high efficiency for ultra-low-bitrate speech coding.
\end{abstract}

\section{Introduction}

A speech codec compresses and reconstructs speech signals to enable efficient transmission and storage \cite{salami1994toll,brandenburg1994iso,salami1997description,keromytis2011comprehensive}. Its core objective is to balance bitrate and reconstruction quality, making speech codecs essential for applications such as real-time communication, voice archiving, and remote conferencing under bandwidth or storage constraints.

With the rapid development of deep learning, neural speech codecs have demonstrated strong rate-distortion performance, achieving a better balance between bitrate and quality than traditional codecs. Waveform-based codecs such as SoundStream \cite{zeghidour2021soundstream} and EnCodec \cite{defossez2023high} directly encode waveforms using causal convolutional networks, while DAC \cite{kumar2024high} further improves fidelity through a non-causal backbone and enhanced quantization. However, waveform-domain modeling can be computationally expensive and may struggle to preserve long-term spectral structure. To address this issue, MDCTCodec \cite{jiang2024mdctcodec} discretizes modified discrete cosine transform (MDCT) spectra to achieve competitive quality with a lightweight architecture.

Despite this progress, a critical challenge remains at ultra-low bitrates (e.g., 0.5~kbps) for scenarios such as satellite communications, on-device storage, and IoT-based voice interfaces, where most neural speech codecs degrade sharply. Most existing codecs rely on residual vector quantization (RVQ) \cite{zeghidour2021soundstream,defossez2023high,kumar2024high,jiang2024mdctcodec}, where reconstruction quality is highly sensitive to the number of quantizers, yet each stage typically consumes the same bitrate, limiting further reduction. Some codecs, such as SQCodec \cite{zhai2025one}, adopt single-codebook finite scalar quantization (FSQ) \cite{mentzer2024finite} to reach lower bitrates, but FSQ is often coarser than VQ and can impair reconstruction quality. Recent approaches such as BigCodec \cite{xin2024bigcodec} and WavTokenizer \cite{jiwavtokenizer} improve ultra-low-bitrate performance by substantially enlarging model capacity, at the cost of heavy models that are less practical for real-world deployment.

To address the ultra-low-bitrate coding challenge, we propose P2PSynCodec, equipped with a novel plain-to-pseudo synergistic vector quantizer (P2PSVQ). Instead of scaling up encoder--decoder capacity, P2PSVQ improves coding efficiency at the quantization level by cascading a plain VQ with multiple pseudo VQs. The plain VQ produces the transmitted tokens, while the pseudo VQs predict auxiliary tokens at the decoder to restore high-bitrate expressiveness without increasing the bitrate. As a result, P2PSynCodec maintains low complexity while significantly improving quality at 0.5~kbps. Experimental results show that P2PSynCodec achieves reconstruction quality comparable to codecs operating at 2.0~kbps, thereby reducing the bitrate by 75\%.

\begin{figure*}
    \centering
    \includegraphics[width=0.9\linewidth]{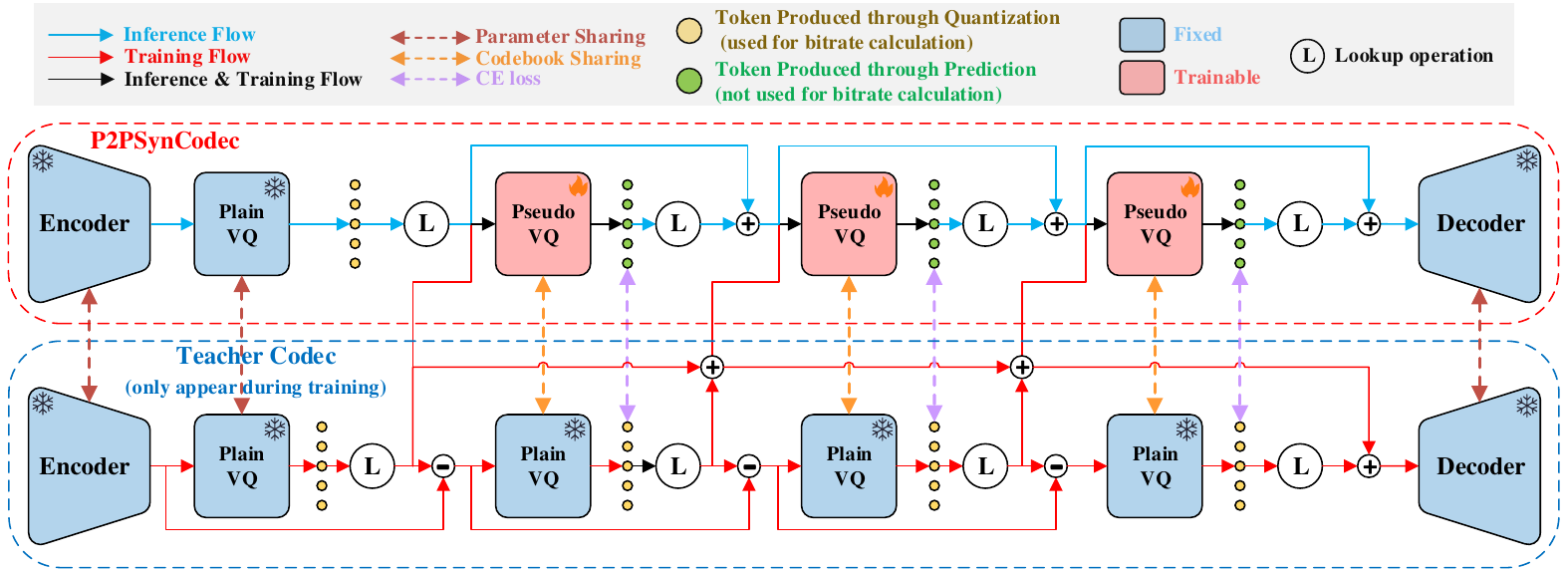}
    \caption{Overview of the proposed P2PSynCodec and its pseudo-VQ training process (illustrated with one plain VQ and three pseudo VQs).
    }
    \label{overview}
\end{figure*}

\section{Proposed Method}
\label{sec:format}
\subsection{Overview}

Fig. \ref{overview} shows an overview of the proposed P2PSynCodec. 
It consists of an encoder, a P2PSVQ, and a decoder, in which the quantizer is a cascaded structure of plain and pseudo VQs.
At the encoding end, the encoder downsamples the input speech to produce compressed encoded representations. 
Subsequently, the P2PSVQ quantizes the continuous encoded representations, in which the plain VQ quantizes basic tokens which can be used for transmission or storage, while the pseudo VQs predict auxiliary tokens based on the basic tokens.
Finally, all tokens are converted into quantized features via codebook lookup, summed, and then fed to the decoder to reconstruct the speech waveform.

\subsection{Encoder \& Decoder}
\label{sec:pagestyle}

Inspired by \cite{jiang2024mdctcodec}, P2PSynCodec operates on MDCT spectra, and implements both the encoder and decoder as fully convolutional networks to keep the model lightweight.
For the encoder, we first extract the MDCT spectra from the input speech and feed them into the network for compression.
The encoder backbone is a modified ConvNeXt v2 \cite{woo2023convnext} network; each residual block consists of a 1D depthwise convolution, layer normalization, a linear layer, global response normalization (GRN), and a GELU activation \cite{hendrycks2016gaussian}.
We additionally place two 1D convolution layers at the input and output of the encoder to adjust feature dimensionality, and use a 1D downsampling layer for temporal compression.
The decoder mirrors the encoder, replacing downsampling with upsampling, and outputs reconstructed MDCT spectra, which are converted back to waveform via inverse MDCT (IMDCT).

\subsection{Plain-to-Pseudo Synergistic Vector Quantizer}

\label{sec:typestyle}
The P2PSVQ in P2PSynCodec consists of a plain VQ and $N$ pseudo VQs. 
Their synergistic design enables P2PSynCodec to achieve ultra-low-bitrate compression while maintaining high reconstruction quality without excessively increasing model complexity. 
Specifically, the plain VQ serves as the foundation, generating basic tokens through quantization. 
The pseudo VQs act as a supplementary module, using neural networks to generate auxiliary tokens through prediction. 
Since these auxiliary tokens are derived from basic tokens rather than obtained via quantization, they do not contribute to bitrate calculation.
Thus, although there are in total $N+1$ VQs, only one plain VQ participates in bitrate calculation, effectively assigning zero bitrate to the pseudo VQs and thereby significantly reducing the overall bitrate.

\subsubsection{Plain VQ}

\label{plain quantizer}

In P2PSVQ, the single plain VQ $Q_{pl}$ discretizes the encoder’s output frame-level vector $\bm{e}\in\mathbb R^K$ through quantization, producing the discrete token $d_{pl}\in\{1,2,\dots,M_{pl} \}$. 
Here, $K$ denotes vector dimension, and $M_{pl}$ is the codebook size of the plain VQ. 
Assume the codebook of the plain VQ is $\mathbb{W}_{pl}=\{\bm{w}_{m}\in\mathbb{R}^K \mid m=1,\dots,M_{pl}\}$, the token $d_{pl}$ is obtained by selecting the index of the code vector with the minimum Euclidean distance to the encoded vector $\bm{e}$, i.e., 
\begin{equation}
d_{pl} = \arg\min_m || \bm{e} - \bm{w}_{m} ||_2.
\end{equation}

Since this single plain VQ directly quantizes the encoded vector, it captures the richest information, and its quantization result is further used by subsequent pseudo VQs to predict auxiliary tokens; therefore, $d_{pl}$ is referred to as the basic token. 
Although P2PSVQ contains $N+1$ VQs, only the token from the single plain VQ is produced through quantization and contributes to bitrate calculation. 
Accordingly, the bitrate of P2PSynCodec is computed as
\begin{equation}
Bitrate = \frac{f_s}{D} \cdot log_2 M_{pl},
\end{equation}
where $f_s$ and $D$ represents the speech sampling rate and the downsampling/upsampling rate of the encoder/decoder, respectively.

\subsubsection{Pseudo VQs}

\label{pseudo quantizer}

The P2PSVQ employs $N$ pseudo VQs (denoted as \(Q_{ps}^{(n)}\), \(n=1,...,N\)), which in practice generate auxiliary tokens through neural network prediction based on the basic token, thereby realizing ``pseudo" quantization. 
The generated auxiliary tokens do not participate in bitrate calculation. 
Each pseudo VQ includes three Conformer blocks \cite{gulati2020conformer} and two bidirectional long short-term memory (BiLSTM) layers \cite{hochreiter1997long}, architectures chosen for their strong ability to capture local spectral patterns as well as long-range temporal dependencies in speech.

As illustrated by the blue and black lines in Fig. \ref{overview}, the process of generating auxiliary tokens is as follows. 
Take the $n$-th pseudo VQ $Q_{ps}^{(n)}$ as an example ($n=1,\dots,N$), it predicts an auxiliary token $\hat{d}_{ps}^{(n)}$ based on the plain token $d_{pl}$ together with the previously generated auxiliary tokens $\hat{d}_{ps}^{(1)},\dots,\hat{d}_{ps}^{(n-1)}$ (if any). 
The pseudo VQ $Q_{ps}^{(n)}$ first performs deep processing on the input using the neural network $NN_{ps}^{(n)}$ (i.e., Conformers + BiLSTMs), producing intermediate feature vector $\bm{z}^{(n)}\in\mathbb R^{M_{ps}^{(n)}}$, i.e.,
\begin{equation}
\scalebox{0.95}{$
\bm{z}^{(n)} =
\begin{cases}
NN_{ps}^{(n)}\left(\mathbb{L}(\mathbb{W}_{pl},d_{pl})\right),&\text{if } n=1,\\
NN_{ps}^{(n)}\left(\mathbb{L}(\mathbb{W}_{pl},d_{pl}) + \sum_{n' = 1}^{n - 1}\mathbb{L}(\mathbb{W}_{ps}^{(n')},\hat{d}_{ps}^{(n')})\right),&\text{if }n>1.
\end{cases}
$}
\end{equation}
where \(\mathbb{L}\) denotes the lookup operation and $\mathbb{W}_{ps}^{(n')}$ denotes the codebook of $Q_{ps}^{(n')}$ with size $M_{ps}^{(n')}$. 
Finally, the auxiliary token \(\hat{d}_{ps}^{(n)}\) is selected as the index of the maximum logit in \(\bm{z}^{(n)}\), i.e.,
\begin{equation}
\hat{d}_{ps}^{(n)} = \arg\max_i z_i^{(n)}.
\end{equation}

\subsubsection{Synergistic Mechanism}

\label{Synergistic Mechanism}

Through a synergistic mechanism, the basic token and auxiliary tokens jointly produce the quantized vector $\hat{\bm{e}}\in\mathbb R^K$ of P2PSVQ, i.e., 
\begin{equation}
\hat{\bm{e}} = \mathbb{L}(\mathbb{W}_{pl},d_{pl}) + \sum_{n= 1}^{N}\mathbb{L}(\mathbb{W}_{ps}^{(n)},\hat{d}_{ps}^{(n)}).
\end{equation} 
Finally, the quantized results are fed into the decoder to reconstruct the speech.

\subsection{Training Paradigm}

\label{training}
The P2PSynCodec adopts a two-stage training strategy.

\subsubsection{Plain-VQ Training Stage}
In this stage, the $N$ pseudo VQs in P2PSVQ are replaced with plain VQs, and then all $N+1$ plain VQs $Q^{(1)}, Q^{(2)},\dots,Q^{(N+1)}$ with codebooks $\mathbb W^{(1)},\mathbb W^{(2)},\dots,$ $\mathbb W^{(n+1)}$ form an RVQ structure, equivalent to MDCTCodec \cite{jiang2024mdctcodec}. 
This codec serves as the teacher model to guide the training in the next stage. 
Its training process adopts the generative adversarial loss, codebook loss and spectral-level loss following \cite{jiang2024mdctcodec}. 

\subsubsection{Pseudo-VQ Training Stage}
At this stage, P2PSynCodec focuses on training the pseudo VQs (with all other modules fixed) under the supervision of the teacher codec, using the teacher-forcing strategy and cross-entropy (CE) loss. 
As illustrated in Fig. \ref{overview}, P2PSynCodec inherits its encoder, plain VQ (i.e., $Q_{pl}=Q^{(1)}$), and decoder entirely from the teacher codec. 
The codebooks of the pseudo VQs in P2PSynCodec are inherited from the corresponding plain VQs of the teacher codec, i.e., $\mathbb W_{ps}^{(n)}=\mathbb W^{(n+1)},n=1,\dots,N$.
For training the pseudo VQs, we adopt the teacher-forcing strategy, as indicated by the red and black lines in Fig.~\ref{overview}. 
Assume that the $N+1$ plain VQs of the teacher codec quantize $\bm{e}$ to produce tokens $d^{(1)}, \dots, d^{(N+1)}$. 
The pseudo VQs are actually trained independently; taking the $n$-th ($n=1,\dots,N$) pseudo VQ as an example, it takes as input the quantization results of the first $n$ plain VQs from the teacher codec and outputs a probability distribution:
\begin{equation}
\tilde{\boldsymbol{p}}^{(n)}=softmax\left[NN_{ps}^{(n)}\left( \sum_{n' = 1}^n\mathbb{L}\left(\mathbb{W}^{(n')},d^{(n')}\right)\right)\right]. 
\end{equation}
On the other hand, the token $\bm{d}^{(n)}$ generates the target probability distribution $\boldsymbol{p}^{(n)}$ through one-hot encoding.
A cross-entropy loss is defined between $\tilde{\boldsymbol{p}}^{(n)}$ and $\boldsymbol{p}^{(n)}$ to minimize the distance between the two distributions, and is used to train the $n$-th pseudo VQ, i.e., 
\begin{equation}
\begin{aligned}
\mathcal{L}^{(n)} =\mathbb{E}_{(\tilde{\boldsymbol{p}}^{(n)}, \boldsymbol{p}^{(n)})} \text{CrossEntropy}(\tilde{\boldsymbol{p}}^{(n)}, \boldsymbol{p}^{(n)}).
\end{aligned}
\end{equation}
The above process is executed from $n=1$ to $N$, completing the training of all $N$ pseudo VQs.

\section{Experiments and Results}

\subsection{Experimental Setup}

\label{ssec:subhead}

Our experiments were conducted on the LibriTTS \cite{zen2019libritts} and VCTK \cite{veaux2017superseded} datasets.
For LibriTTS, with a sampling rate of 16 kHz, the training process utilized the train-clean-100 and train-clean-360 subsets, while the dev-clean and test-clean subsets were employed for validation and evaluation, respectively.
As for VCTK, with a sampling rate of 48 kHz, its training set consisted of 40,936 utterances, and the test set was made up of 2,937 utterances.

In P2PSynCodec\footnote{Speech samples can be accessed at: 
\href{https://pb20000090.github.io/P2PSynCodec/}{https://pb20000090.github.io/P2PSynCodec/}.}, the P2PSVQ employed three pseudo VQs (i.e., $N=3$), all with the same codebook size of 1024 (i.e., $M_{pl}=M_{ps}^{(1)}=M_{ps}^{(2)}=M_{ps}^{(3)}=1024$) and code vector dimension of 32 (i.e., $K=32$).
In each pseudo VQ, the Conformer block had 256 channels and 8 attention heads, while the BiLSTM had 256 channels.
The downsampling/upsampling rate was set to $D=320$. 
Therefore, the bitrate of P2PSynCodec is just 0.5 kbps for 16 kHz sampling rate (i.e., $f_s=16000$) and 1.5 kbps for 48 kHz sampling rate (i.e., $f_s=48000$).

\subsection{Evaluation Metrics}

\label{sssec:subsubhead}

For objective evaluation, we adopted both non-intrusive and intrusive metrics. The non-intrusive metrics UTMOS \cite{saeki2022utmos} and SIGMOS \cite{ristea2025icassp} were used at 16~kHz and 48~kHz, respectively, to assess overall speech quality. We further included STOI \cite{taal2010short} and ViSQOL \cite{chinen2020visqol} as intrusive metrics to measure intelligibility and reference-based quality.
In addition, floating point operations (FLOPs) \cite{mcmahon1986livermore} and the number of parameters (Param.) were used to evaluate computational complexity and model complexity, respectively.

For subjective evaluation, we conducted multiple stimuli with hidden reference and anchor (MUSHRA) \cite{recommendation2001method} and ABX preference tests on Amazon Mechanical Turk to compare P2PSynCodec with the baseline codecs.
For MUSHRA, 20 test-set utterances per codec were randomly selected from the test set and rated by at least 25 native English listeners on a 0--100 scale, with natural speech as the hidden reference and a 3.5-kHz low-pass-filtered version as the anchor.
For ABX, 20 test-set utterance pairs for each comparison were evaluated by at least 25 native English listeners, who were asked to determine which utterance in each pair had better speech quality, or whether they had no preference.
We report mean scores/preferences and assess significance using a $t$-test ($p$-value).

\subsection{Comparison with Baseline Neural Speech Codecs}
\label{sec:print}

We compared the proposed P2PSynCodec with several advanced baseline neural speech codecs, including RVQ-based MDCTCodec \cite{jiang2024mdctcodec} and DAC \cite{kumar2024high}, single-VQ-based BigCodec \cite{xin2024bigcodec} and WavTokenizer \cite{jiwavtokenizer}, and FSQ-based SQCodec \cite{zhai2025one}.

\subsubsection{Comparisons at Equal Ultra-Low Bitrates}
\label{sec:comparisons}
For a fair comparison, all baselines were configured to match the ultra-low bitrates of P2PSynCodec (0.5~kbps at 16~kHz and 1.5~kbps at 48~kHz). Specifically, we matched the target bitrate by setting the overall downsampling factor to 320 (with stride factors of 2, 4, 5, and 8), using a single codebook, and setting the codebook size to 1024. This results in 50 tokens per second at 16~kHz and 150 tokens per second at 48~kHz, corresponding to target bitrates of 0.5~kbps and 1.5~kbps, respectively.
SQCodec was excluded since its official release does not support this setting.

\begin{figure}[t]
  \centering
  \includegraphics[width=0.9\linewidth]{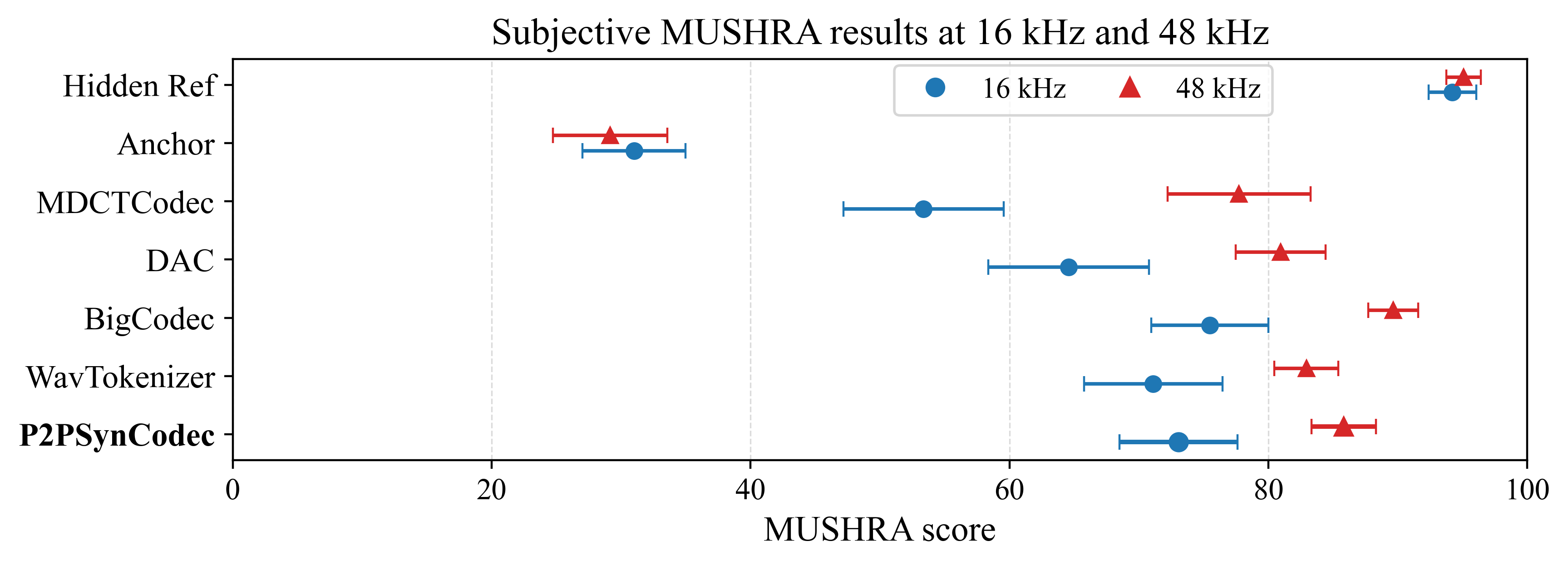}
 \caption{Subjective MUSHRA results at 16 and 48 kHz, including the hidden reference and anchor. Error bars denote 95\% confidence intervals.}
  \label{mushra}
\end{figure}

Table~\ref{16k} summarizes the results. We first compare P2PSynCodec with the RVQ-based codecs MDCTCodec and DAC. On the overall-quality metrics, P2PSynCodec surpasses both baselines by more than one point on UTMOS at 16~kHz and remains clearly superior on SIGMOS at 48~kHz, indicating a consistent perceptual advantage in both settings. Although P2PSynCodec does not always achieve the best scores on intrusive metrics, this is expected because generative codecs are inherently disadvantaged under reference-based evaluation, since metrics such as PESQ \cite{rix2001perceptual}, POLQA \cite{beerends2013perceptual}, and ViSQOL \cite{chinen2020visqol} \emph{cannot} always accurately evaluate generative models like our pseudo VQs, as also noted in \cite{maiti2020speaker,yao2025gense,hsu2023revise}. To provide further evidence, the subjective MUSHRA results in Fig.~\ref{mushra} show that P2PSynCodec also achieves higher scores than both MDCTCodec and DAC, with a particularly large margin over MDCTCodec. 

We then compare P2PSynCodec with the single-codebook codecs WavTokenizer and BigCodec. P2PSynCodec consistently outperforms WavTokenizer in terms of both the objective metrics and the MUSHRA scores, suggesting that the proposed plain-to-pseudo synergistic quantization more effectively recovers expressiveness under the same bitrate budget. P2PSynCodec is also comparable to BigCodec in reconstruction quality, as measured by UTMOS, SIGMOS, and MUSHRA, while using only about 5\% of its FLOPs and 14\% of its parameters.

\subsubsection{Comparisons with High-Bitrate Codecs} 
\begin{figure}[t]
  \centering
  \includegraphics[width=0.9\linewidth]{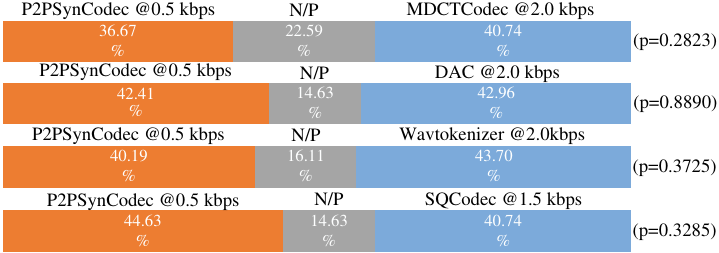}
  \caption{
Average preference scores (\%) of ABX tests comparing P2PSynCodec at 0.5 kbps and other codecs at high bitrates on the LibriTTS test set (16 kHz). Here, N/P denotes ``no preference”, and $p$ is the paired $t$-test $p$-value.
  }
  \label{ABX_1}
\end{figure}

To quantify the bitrate savings brought by the P2PSVQ strategy in P2PSynCodec, we conducted subjective ABX evaluations on LibriTTS (16 kHz), comparing P2PSynCodec at 0.5 kbps with other codecs operating at higher bitrates, including MDCTCodec, DAC, WavTokenizer at 2 kbps, and SQCodec at 1.5 kbps. 
BigCodec was excluded, as its objective results at 0.5 kbps were already comparable to ours. 
As shown in Fig.~\ref{ABX_1}, P2PSynCodec at 0.5 kbps achieved perceptual speech quality comparable to all baseline codecs operating at higher bitrates ($p>0.01$). 
In particular, compared with MDCTCodec at 2 kbps, which serves as our teacher codec, this result validates the effectiveness of trading a moderate increase in model complexity for substantial bitrate savings.
The above results confirm that by introducing the P2PSVQ strategy, P2PSynCodec achieves a 75\% reduction in bitrate.

\begin{table}[t]
\centering
\huge
    \caption{Objective experimental results on decoded speech quality and complexity of the compared codecs at 0.5 kbps on the LibriTTS test set (16 kHz) and 1.5 kbps on the VCTK test set (48 kHz). The \textbf{bold} and \underline{underlined} numbers indicate optimal and sub-optimal results, respectively.}
    \resizebox{\linewidth}{!}{
    \begin{tabular}{c | c c c | c c c| c c}
        \hline
        
        \hline
        
        \hline
        &\multicolumn{3}{c|}{LibriTTS (16 kHz, 0.5 kbps)}&\multicolumn{3}{c|}{VCTK (48 kHz, 1.5 kbps)}&\multirow{2}{*}{FLOPs} &\multirow{2}{*}{Param.}\\
        \cline{2-7}
        &UTMOS &STOI &ViSQOL & SIGMOS & STOI & ViSQOL & \\ 
         \hline
         MDCTCodec & 2.670 & \underline{0.844} & \underline{3.631} & 2.846  & \underline{0.828} & \textbf{3.673} &\textbf{2.32G}& \textbf{6.75M} \\
         DAC & 2.725 & 0.818 & 3.386 & 2.971 & 0.790 & 3.425 &55.53G& 73.87M\\
         BigCodec &\underline{3.939}& \textbf{0.872} & \textbf{3.682} & \underline{3.277} & \textbf{0.840} & \underline{3.478} &61.03G& 159.32M\\
         
         WavTokenizer & 3.269 & 0.834 & 3.484 & 3.232 & 0.784 & 3.255 & 4.21G & 71.65M \\
         P2PSynCodec&\textbf{3.947}& 0.823 & 3.476 & \textbf{3.305} & 0.796 & 3.423 &\underline{3.31G}&\underline{22.99M}  \\
        \hline
        
        \hline
        
        \hline
    \end{tabular}}
\label{16k}
\end{table}

\begin{table}[t]
\centering
    \caption{Objective experimental results of the analysis experiment on the number of pseudo VQs on the LibriTTS test set (16 kHz).}
    \resizebox{0.9\linewidth}{!}{
    \begin{tabular}{c | c c c| c c c}
        \hline

        \hline

        Number of &\multicolumn{3}{c|}{All VQs}&\multicolumn{3}{c}{Only Plain VQ}\\
         \cline{2-7}
         Pseudo VQs
       
         &UTMOS &STOI & ViSQOL &UTMOS &  STOI & ViSQOL\\ 
         \hline
         
         1 & 3.787 & 0.845 & 3.551 & 3.048 & 0.840 & 3.581\\
         3 & 3.947 & 0.823 & 3.476 & 2.324 & 0.806 & 3.498\\
         5 & 3.986 & 0.798 & 3.208 & 1.943 & 0.756 & 3.229\\
         7 & 3.889 & 0.725 & 2.761 & 1.296 & 0.699 & 2.775 \\
        \hline

        \hline
    \end{tabular}}
\label{abla}
\end{table}

\subsection{Analysis of the Impact of Pseudo VQ Number}

\label{sec:page} 
In this section, we explore the impact of the pseudo VQ number $N$ in P2PSynCodec, which determines both the quality of reconstructed speech and the model complexity. 
The experiments were conducted on LibriTTS, and the objective results are shown in Table \ref{abla}. 
We set $N$ to 1, 3, 5, and 7, and evaluated not only the metrics of the final reconstructed speech (i.e., All VQs in Table \ref{abla}) but also those of the speech decoded using only the quantization result of the plain VQ (i.e., Only Plain VQ in Table \ref{abla}), in order to analyze their relationship.


Interestingly, as shown in Table~\ref{abla}, the decoded speech quality of P2PSynCodec is not positively correlated with the number of pseudo VQs. 
When $N$ becomes large, the objective metrics instead decrease. 
In contrast, the quality of speech decoded from the plain VQ’s quantization results deteriorates markedly with increasing $N$. 
However, this degradation is not well reflected by ViSQOL: its scores remain nearly unchanged and are even higher in some cases than those obtained with all VQs. 
By contrast, UTMOS more faithfully reflects the change in overall speech quality. 
This observation is also consistent with the limitation of intrusive metrics mentioned in Section~\ref{sec:comparisons}. 
This suggests that although increasing $N$ improves the overall decoding quality of the teacher codec (i.e., the upper performance bound of P2PSynCodec), the first plain VQ carries progressively less information, as its load is increasingly distributed across the other VQs. 
Therefore, when $N$ becomes large, the pseudo VQs must predict auxiliary tokens from basic tokens that carry less information, making prediction more difficult and leading to degraded quality. 
In this situation, the ``basic'' and ``auxiliary'' tokens gradually lose their intended roles. 
Moreover, a larger $N$ also results in higher model complexity. 
When $N$ is small, although the plain VQ quantizes sufficient information, the decoding quality of the teacher codec is low, which limits the upper performance bound of P2PSynCodec and results in unsatisfactory decoded speech quality. 
Therefore, the choice of $N$ should be moderate. The current setting of $N=3$ represents an optimal trade-off between decoded speech quality and model complexity.

\section{Conclusion}
In this paper, we proposed P2PSynCodec, an ultra-low-bitrate neural speech codec with a plain-to-pseudo synergistic vector quantizer (P2PSVQ).
The plain VQ generates the transmitted tokens, while pseudo VQs predict auxiliary tokens to enrich the representation without increasing bitrate.
Trained with teacher forcing using an RVQ-based teacher codec, P2PSynCodec preserves expressiveness at ultra-low bitrate with lightweight complexity.
Experiments show that P2PSynCodec at 0.5~kbps achieves speech quality comparable to 2.0~kbps codecs.
Future work will extend the framework to causal architectures for real-time and streaming applications.

\section{Acknowledgments}
This work was supported by the National Natural Science Foundation of China under Grant No. 62301521.

\section{Generative AI Use Disclosure}
During the preparation of this manuscript, the authors used ChatGPT 5.2 to polish the language and improve the flow of the text. After using this tool, the authors reviewed and edited the content as needed and take full responsibility for the final version of the manuscript.

\bibliographystyle{IEEEtran}
\bibliography{mybib}

\end{document}